\documentclass[aps,floatfix,twocolumn,footinbib]{revtex4}
\usepackage{graphicx,color}
\usepackage{amsmath,amssymb}
\usepackage{subfigure}
\usepackage{bm}
\begin{document}

\title{Magnetoelectric bistabilities in ferromagnetic resonant tunneling structures}

\author{Christian Ertler\footnote{email:christian.ertler@physik.uni-regensburg.de}}
\affiliation{Institute for Theoretical Physics, University of
Regensburg, Universit\"atsstrasse 31, D-93040 Regensburg, Germany}

\begin{abstract}

The conditions for the occurrence of pronounced magnetoelectric
bistabilities in the resonant tunneling through a ferromagnetic
quantum well are theoretically investigated. The bistability appears
due to the mutual feedback of the carriers Coulomb interaction and
the carriers exchange coupling with magnetic impurities in the well.
It is shown that the well Curie temperature depends strongly on the
relative alignment of the quantum well level and the reservoirs
chemical potentials, which can be modified electrically. Switching
between a ''current-on/magnetism-off`` and a
''current-off/magnetism-on`` mode becomes possible, if the well
temperature lies in-between the bistable values of the well Curie
temperature.

\end{abstract}

\maketitle

In ultimate magnetoelectric devices the magnetic properties should
be ideally controllable to a vast extent by external bias or gate
fields. For this purpose, band-engineered magnetic resonant tunneling
structures are very promising, since they exhibit a
rich variety of tunable magneto-transport properties
\cite{Fabian2007:APS}. Especially, the impetuous development of
novel dilute magnetic semiconductors (DMSs)
\cite{Ohno1998:S,Dietl:2007,Jungwirth2006:RMP} in the last decades,
which are made magnetic by randomly doping with transition metal elements, e.g., by
incorporating Mn in a GaAs crystal host,
has considerably enriched the possibilities of growing different
magnetic semiconductor heterostructure systems. In DMSs the ferromagnetism can
depend strongly on the actual particle density, which has been
confirmed in several experiments, in which ferromagnetism
has been generated by tailoring the particle density by electrical or optical means \cite{Ohno2000:N, Boukari2002:PRL}.

In magnetic resonant tunneling structures made of para- or
ferromagnetic DMSs even small energetic spin splittings of the well
subbands can become observable in the transport characteristics
\cite{Slobodskyy2003:PRL,Ohya2007:PRB, Oiwa2004:JMMM}. Based on
their spin-dependent transmission magnetic resonant tunneling
structures have been proposed for realizing efficient spin valves
and spin filtering devices \cite{Fabian2007:APS, Petukhov2002:PRL},
or for digital magnetoresistance \cite{Ertler2007a:PRB,
Ertler2006a:APL}. The magnetic properties of ferromagnetic quantum
wells made of DMSs are well described in the framework of a mean
field model \cite{Dietl1997:PRB,Jungwirth1999:PRB, Lee2002:SST,
Ganguly2005:PRB}, which reveals that the Curie temperature of the
well depends on (i) the 2D-spin susceptibility of the carriers and
(ii) on the overlap of the subband wave function with the magnetic
impurity density profile. Both parameters should be in principle
tuneable by the applied bias, which would provide a purely {\em
electrical} control of the ferromagnetism in magnetic quantum wells.

In conventional nonmagnetic resonant tunneling diodes (RTDs) it is
well known, that an intrinsic hysteresis in the
negative-differential-resistance (NDR) region of the current-voltage
(IV) characteristics can occur \cite{Goldman1987:PRL}. This
bistability of the tunneling current has been explained to result
from the nonlinear feedback of Coulomb interaction of the stored
well charge \cite{Sheard1988:APL,Kluksdahl1989:PRB}. In magnetic
RTDs this naturally suggests the possibility of hysteretic magnetic
states, as has been predicted in
\cite{Sanchez2001:PRB,Ganguly2006b:PRB}.

In this article a detailed study of possible magnetoelectric
bistabilities in magnetic RTDs is provided. The carriers dynamics is
described by a self-consistent sequential tunneling model, which
includes the feedback effects of both the carriers Coulomb
interaction and the magnetic exchange coupling with the magnetic
ions. The model yields a simple expression for the steady state
2D-spin susceptibility, which allows to calculate the critical
temperature $T_c$ of the quantum well depending on the applied bias
and the relative alignment of the quantum well level with respect to
the chemical potentials of the emitter and collector reservoirs. If
the well is operated at a temperature, which lies between the
bistable values of the well Curie temperature, the magnetic RTD can
be switched between a ''current-on/magnetism-off`` and a
''current-off/magnetism-on`` mode.

\begin{figure}
\centerline{\includegraphics{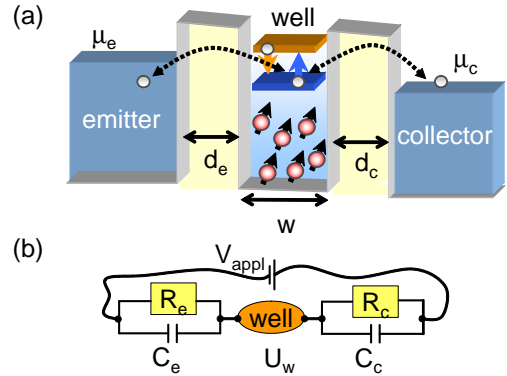}}
\caption{(Color online) (a) Schematic scheme of the band profile of
the magnetic double barrier structure. The exchange interaction of
the magnetic ions is mediated by the carriers tunneling in and out
of the well. (b) Equivalent circuit model of
the resonant tunneling structure introducing the emitter and
collector capacitances $C_e, C_c$ and resistances $R_e, R_c$,
respectively.} \label{fig:structure}
\end{figure}

The band profile of a generic double-barrier resonant tunneling
structure with a ferromagnetic quantum well made of a DMS, e.g., of
GaMnAs, is sketched in Fig.~\ref{fig:structure}(a). The vertical
transport through the structure can be described by a sequential
tunneling model, since the high density of magnetic impurities in
the well will likely cause decoherence processes. By using the
transfer Hamiltonian formalism a Pauli master equation for the
statistical distribution of the particles in the well can be derived
\cite{Fabian2007:APS, Averin1991:PRB}. In the case that only a
single resonant level $E_w$ resides in the energy window of
interest, which is defined by the difference of the emitter's and
collector's chemical potentials, simple rate equations for the
spin-resolved well particle densities $N_\sigma(t)$ with $(\sigma =
\uparrow,\downarrow)$ are obtained
\begin{equation}\label{eq:rate}
\frac{\mathrm{d}N_\sigma}{\mathrm{d} t} = \Gamma_{e}(E_\sigma)\:
N_{e,\sigma} + \Gamma_{c}(E_\sigma)\:
N_{c,\sigma} - \Gamma(E_\sigma)\: N_\sigma.
\end{equation}
Here,  $N_{\sigma,\{e,c\}}$ are
the densities of particles with the resonant longitudinal energy $E_\sigma$
in the emitter (e) and collector (c) reservoir,
respectively. The energy-dependent tunneling rates $\Gamma_{\{e,c\}},
\Gamma = \Gamma_e+\Gamma_c $ can be calculated by Bardeen's
formula \cite{Bardeen1961:PRL}, which essentially evaluates  the
overlap of the lead and well wave functions in the barriers.
For high barriers these tunneling rates become proportional to the
longitudinal momentum $p_z$
of the incident particles \cite{Averin1991:PRB}, i.e., $\Gamma_{e,c} \propto (E_z)^{1/2}$ with
$E_z$ denoting the longitudinal energy.
By assuming that the particle reservoirs are described by Fermi-Dirac
distributions the particle densities are given by
$ N_{i,\sigma} = D_0 k_B T
\ln\left\{1+\exp\left[(\mu_i-E_\sigma)/k_B T\right]\right\}$,
$i=(e,c),$
with  $D_0 = m/2\pi\hbar^2 $ is
the two-dimensional density of states per spin for carriers with the
effective mass $m$,
$k_B$ denotes  Boltzmanns' constant, $T$ is the lead
temperature, and $\mu_i$ are the emitter and collector chemical
potentials with $\mu_c = \mu_e-e V_\mathrm{appl}$ where
$V_\mathrm{appl}$ is the applied bias.

In the framework of a mean field model an
analytic expression for the steady state exchange splitting $\Delta$ of the
well level can be derived
\cite{Dietl1997:PRB,Jungwirth1999:PRB, Lee2002:SST, Fabian2007:APS}
\begin{eqnarray}
\Delta &=& J_\mathrm{pd}\int\mathrm{d}z\:
n_\mathrm{imp}(z)\left|\psi_0(z)\right|^2\nonumber\\
\label{eq:exchange}&& \times S B_S\left[\frac{S J_\mathrm{pd} s
(N_\downarrow- N_\uparrow) \left|\psi_0(z)\right|^2}{k_B T}\right],
\end{eqnarray}
where $J_\mathrm{pd}$  is the coupling strength between the
impurity spin and the carrier spin density (in case of GaMnAs p-like
holes couple to the d-like impurity electrons), $z$ is the
longitudinal (growth) direction of the structure,
$n_\mathrm{imp}(z)$ is the impurity density profile,  $\psi_0(z)$
labels the well wave function, and $s=1/2$ is the particles spin. The Brillouin function
of order $S$ is denoted by $B_S$, where $S $ is the impurity spin,
which for Mn equals 5/2.
By considering a homogenous impurity distribution $\Delta$ is effectively
determined by the voltage dependent spin polarization $\xi = s(N_\uparrow- N_\downarrow)$.

The nonlinear feedback of the Coulomb interaction of the well
charges is approximately taken into account by calculating the
electrostatic well potential in terms of an equivalent circuit model
of the resonant tunneling diode, as shown in
Fig.~\ref{fig:structure}(b), where the capacitances $C_e$ and $C_c$
are determined by the geometrical dimensions of the barriers and the
well \cite{Averin1991:PRB}. The potential results in $U_w =
[e^2(N-N_\mathrm{back})-C_c eV_\mathrm{appl}]/C$, where $N =
N_\uparrow+N_\downarrow$, $C = C_e+C_c$, $e$ denotes the elementary
charge, and $N_\mathrm{back}$ is the positive background charge in
the well, which originates from the magnetic donors. Since the
actual position of the quantum well levels $E_\sigma =
E_0+U_w-\sigma\Delta$ depends on both the magnetic exchange
splitting $\Delta$ and the electrostatic potential $U_w$ all
equations become nonlinearly coupled, making a selfconsistent
numerical solution necessary.

\begin{figure}
\centerline{\includegraphics{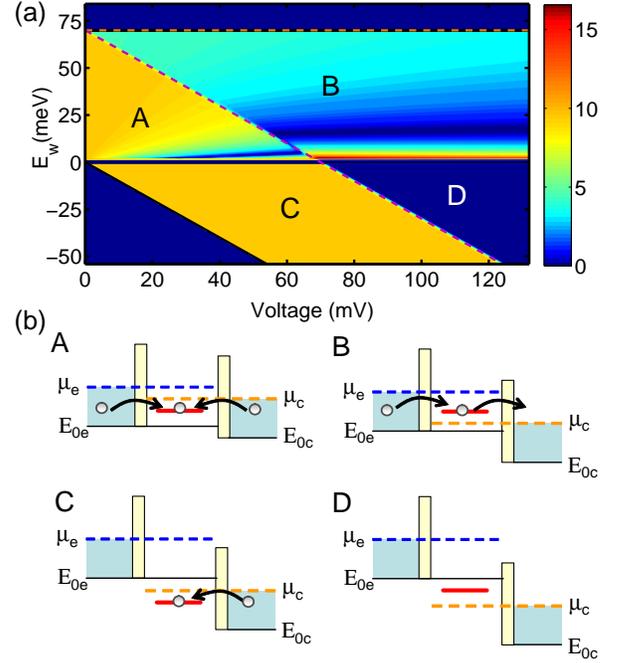}} \caption{(Color online) (a)
Contour plot of the well Curie temperature $T_c$ (K) as a function
of the well level position $E_w$ and the applied bias. (b) Schematic
illustration of the different occupation probabilities of the
quantum well level from the reservoirs for the regions A-D, as
indicated in the contour plot (a). The emitter and collector band
edges are denoted by $E_{0e}$ and $E_{0c}$, respectively.}
\label{fig:phdiag}
\end{figure}

In order to find criterions for the occurrence of magnetic
bistabilities and to interpret the numerical results in the
following it is very useful to study the dependence of the well
Curie temperature $T_c$ on the applied bias and the well level
position. The mean field model yields an analytic expression for the
collective Curie temperature of a magnetic quantum well
\begin{equation}
k_B T_c = \frac{S(S+1)}{3}J_{\mathrm{pd}}^2 \chi_{2D}\int\mathrm{d} z\:
 n_i(z) |\psi(z)|^4,
\end{equation}
where the two-dimensional spin susceptibility is defined by
\begin{equation}
\chi_{2D} = \lim_{\Delta \rightarrow 0}\frac{s
(N_\uparrow-N_\downarrow)}{E_\downarrow-E_\uparrow}.
\end{equation}
Within the introduced sequential tunneling model Eq.~(\ref{eq:rate})
the steady state spin susceptibility simplifies to $\chi_{2D}(E) =
-s (\partial N_0/\partial E)$, where $N_0 = (\Gamma_e N_e + \Gamma_c
N_c)/\Gamma$ is the steady state solution of the rate equations
(\ref{eq:rate}). Hence, the dimensionless susceptibility can be
written as
\begin{equation}\label{eq:chis}
\tilde{\chi}=\frac{\chi_{2D}(E)}{s D_0} = \sum_{i=e,c}\frac{\Gamma_i}{\Gamma}f_{\mathrm{FD}}^i
-\frac{N_i}{D_0}
\frac{\partial}{\partial E}\left(\frac{\Gamma_i}{\Gamma}\right)
\end{equation}
with $f_{\mathrm{FD}}^i, i=(e,c)$ denoting the Fermi-Dirac function
for the emitter and collector reservoir, respectively. This allows
to calculate the Curie temperature $T_c$ as a function of the
applied voltage (note that $T_c$ depends via $f_\mathrm{FD}^c$
explicitly on the voltage, since $\mu_c = \mu_e - e
V_{\mathrm{appl}}$) and the well level position, as displayed in
Fig.~\ref{fig:phdiag}(a).
 For the simulations I used
generic parameters corresponding to a GaMnAs well: $m = 0.5\:m_0$,
$\varepsilon_r = 12.9$, $d_e = d_c= 20$ \AA, $w = 10$ \AA, $\mu_e =
70$ meV, $n_\mathrm{imp} = 1.5\times10^{20} $cm$^{-3}$,
$J_{\mathrm{pd}} = 0.06$ eV nm$^3$, where $d_e, d_c$ and $w$ are the
emitter barrier, collector barrier and quantum well widths, $m_0$
denotes the free electron mass, and $\varepsilon_r$ is the relative
permittivity of the well. The background charge $n_\mathrm{back} =
0.1\:n_\mathrm{imp}$ is considered to be only of about 10\% of the
nominal Mn doping density \cite{DasSarma2003:PRB} and the lattice
temperature is set to
 $T = 4.2$~K.

\begin{figure}
\centerline{\includegraphics{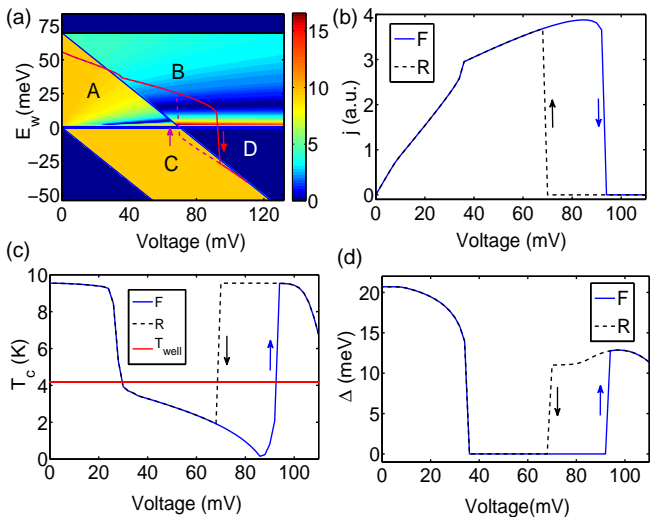}} \caption{(Color online) The
(a) quantum well level position $E_w$, (b) current $j$, (c) Curie
temperature $T_c$, and (d) well splitting $\Delta$ as a function of
the applied bias. The solid lines indicate the voltage up-sweep
values (F), whereas the dashed lines correspond to the voltage
down-sweep values (R). In (a) the $E_w$-voltage curves are embedded
in the contour plot of the well Curie temperature $T_c$ (K) and in
(d) the solid red line corresponds to the actual well temperature
$T_{\mathrm{well}} = 4.2$~K. \label{fig:bistab}}
\end{figure}

The Curie temperature contour plot can be divided into four
qualitatively different regions A-D, which are characterized by
different probabilities for occupying the quantum well level from
the reservoirs, as schematically illustrated in
Fig.~\ref{fig:phdiag}(b). In region A, for instance, the well level
can be occupied by particles originating from both reservoirs. The
Curie temperatures $T_c^{(A,C)}$ in regions A and C differ roughly
by a factor 2 compared to $T_c^{B}$ of region B. This sudden change
of $T_c$ can be explained as follows: by assuming energy-independent
tunneling rates and nearly symmetric barriers, i.e., $\Gamma_e
\approx \Gamma_c$  the dimensionless spin susceptibility of
Eq.~(\ref{eq:chis}) simplifies for region A to $\tilde{\chi}^{A} =
1/2(f_\mathrm{FD}^e+f_\mathrm{FD}^2) \approx 1$, whereas in the
other regions one obtains: $\tilde{\chi}^{B} = f_\mathrm{FD}^e/2
\approx 1/2, \tilde{\chi}^{C} = f_\mathrm{FD}^c \approx 1$, and
$\tilde{\chi}^{D} = 0$. This simple estimation, hence, yields the
desired result $T_c^{(A,C)}/T_c^{B} \approx 2$.

These differences in the Curie-temperatures of the various regions
can now be exploited to realize hysteretic magnetoelectric states.
According to the nonlinear feedback of the stored well charge, the
resonant level $E_w$ and the IV-characteristic show a hysteretic
behavior, as displayed in Fig.~\ref{fig:bistab}(a) and (b). For the
up-sweep (F) of the applied bias the well is charged before the
$E_w$ becomes off-resonant, i.e., when it drops below the emitter
band edge, whereas for the voltage down-sweep (R) the well is almost
uncharged before $E_w$ becomes resonant again. This leads to
different self-consistent electrostatic potentials for up- and
down-sweeping voltages, explaining the occurrence of the intrinsic
bistability. If the hysteresis of $E_w$ now switches exactly between
the $T_c$-regions B and C, as it is the case in
Fig.~\ref{fig:bistab}(a), then also the voltage-dependent Curie
temperature will exhibit a pronounced hysteresis, as shown in
Fig.~\ref{fig:bistab}(c). The electric hysteresis will then be
accompanied by a magnetic hysteresis if the actual lattice
temperature $T$ of the quantum well, which is displayed as straight
solid line in Fig.~\ref{fig:bistab}(c), fulfills the condition
$T_c^{B} < T < T_c^{C}$. This is illustrated in
Fig.~\ref{fig:bistab}(d): as long as the resonant level stays in
region B the well is nonmagnetic ($\Delta = 0$, since $T > T_c^B$)
but when $E_w$ enters the region C the well becomes immediately
magnetic ($\Delta \neq 0$). Also notice, that at low voltages the
well is always magnetic. At roughly $V_\mathrm{appl}\approx 30$~mV
the well becomes nonmagnetic, since $E_w$ crosses the boundary
between the regions A and B, which provides a purely electrical
control of the well magnetism. As a whole, the magnetic well
switches between a ''current-on/magnetism-off`` state for the
up-sweep and a ''current-off/magnetism-on``-state for the down-sweep
of the applied voltage. Moreover, the $T_c$ contour plot in
Fig.~\ref{fig:phdiag}(a) also suggest the possibility for realizing
the switching between a ''current-on/magnetism-on`` and a
''current-off/magnetism-off'' mode in the case that the hysteresis
of $E_w$ switches between region B and D and if $0<T<T_c^{B}$.

In summary, I have shown by using a selfconsistent sequential
tunneling model that the Curie temperature of a magnetic quantum
wells strongly depends on the relative alignment of the well level
and the reservoirs chemical potentials, which can be modified by
external bias or gate fields. Magnetoelectric bistabilities become
possible if the hysteresis of the well level position $E_w$ switches
between regions of different well Curie temperatures and if the
actual well temperature lies in-between these two values.

This work has been supported by the DFG SFB 689. The author thanks
J. Fabian for very valuable and inspiring discussions.

\end{document}